# Integrated Structured Light Architectures

Randy Lemons[1,2], Wei Liu[1], Josef C. Frisch[1], Alan Fry[1], Joseph Robinson[1], Steve Smith[1], Sergio Carbajo[1]*

[1]SLAC National Accelerator Laboratory and Stanford University, 2575 Sand Hill Road, Menlo Park, CA 94025, USA
[2]Department of Physics, Colorado School of Mines, Golden, CO 80401, USA
*corresponding author: scarbajo@stanford.edu

The structural versatility of light underpins an outstanding collection of optical phenomena where both geometrical and topological states of light can dictate how matter will respond or display. Light possesses multiple degrees of freedom such as amplitude, and linear, spin angular, and orbital angular momenta[1], but the ability to adaptively engineer the spatio-temporal distribution of all these characteristics is primarily curtailed by technologies used to impose any desired structure to light[2]. We describe a foundational demonstration that examines a laser architecture offering integrated spatio-temporal field control and programmability, thereby presenting unique opportunities for generating light by design to exploit its topology.

Structured photonics lay the foundation for the generation or use of light with custom spatio-temporal variant field vector, amplitude and phase distribution. Coloration in peacock feathers and photonic band structures in butterfly wings are among the many complex morphologies of light commonly found in nature. In recent decades, artificial structuring of light has undergone a remarkable evolution to produce orbital and spin angular momenta beams exhibiting unique properties such as optical vortices and topological vector fields. Beyond areas of recent decadal impact, such as optical communications[3], sensing[4], and particle trapping[5], today these properties are examined to create transformational tools in molecular physics[6–8], quantum[9], relativistic[10,11], and nonlinear optics[12], and particle physics[13], to name a few. Unconventional ways of conceptualizing light structure are inspiring new families of electromagnetic fields that circumvent generally unquestioned behavioral properties. For example, Bessel-Bessel-Bessel light bullets[14] can propagate without appreciable diffraction distortion, while an ensemble of finite-energy wavepackets are capable of abruptly focusing and defocusing outside the paradigm of paraxial optics[15]. As quanta, spatio-temporally variant topological states of light, particularly if they can be changed dynamically, could enable photonic and information technologies in quantum computation and highly-selective light-matter interactions, including Floquet insulators[16], photonic skyrmions[17] or two-dimensional metasurface polaritons[18]. In essence, the outlook of structured photonics applications is extraordinary and has no end in sight, but the fruition of this promise is severely hampered by bottleneck technologies that have failed to advance the generation of light with adaptable structure.

One common way of engineering structured light is by using spatial light modulators. These devices can control the intensity and phase of a light beam in image or Fourier space. They represent the success story of light shaping technologies and their widespread application in holographic display technology and optical tweezers, for example. But expanding on this success can prove extremely difficult if it is to depend solely on development of these external modulators. Important parameters that also define light structure can often be left out, such as temporal intensity distribution of light bullets or pistons, or active control of the carrier-envelope phase. But perhaps their main limitation is their operational damage threshold, thereby impeding progress on structured light applications where moderate to extremely high peak- or average-power levels are at play, above MW- and W-levels, respectively[19]. These constraints motivate an alternative approach which is power scalable and where the parameter space that defines structure—transverse and longitudinal wavevector distribution, amplitude and phase—is programmable and decoupled.

In this letter, we present a generalized laser architecture and experimental demonstration that enables the design of light bullets with built-in programmable structure to be exploited adaptively. We refer to this architecture as the Universal Light Modulator (ULM). It capitalizes on advanced composite pulse synthesis designs which have long sought to top its ever-increasing power and intensity to introduce programmable control over the spatio-temporal intensity and field-amplitude, -phase, and -polarization distribution. The underlying principle of the ULM exploits a number of optically coherent frequency and spatial combs, here referred to as beamlines, where each undergoes both intra-beamline (independent) and inter-beamline (self-referenced) manipulation of the electromagnetic fields with high-fidelity control. Fig. 1.a illustrates this principle, where the primary field properties can be controlled in a spatial and temporal optical comb. When a coherent relationship is maintained between all the beamlines, these combs can be made to collapse and generate unique spatio-temporal wavevector distributions.

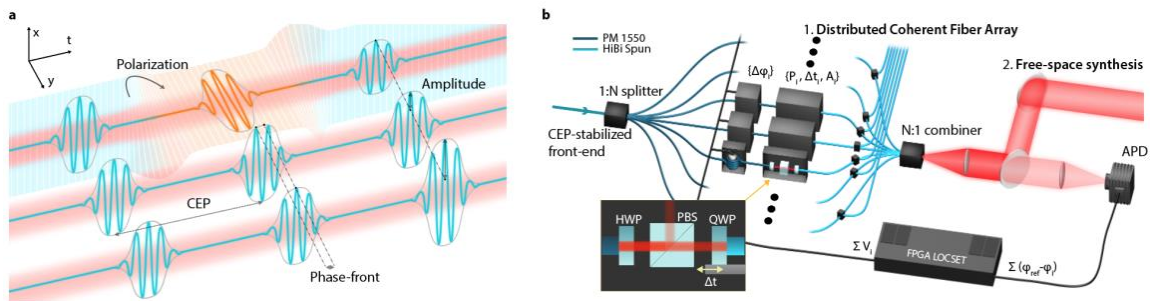

Fig. 1—(a) conceptual depiction of a coherent transverse and longitudinal comb and (b) its experimental configuration via coherent multi-channel fiber-based array

The proof-of-concept ULM consists of N=7+1 ($i = 1:N$) fiber-based beamlines, each split from a femtosecond mode-locked laser operating at the C-band telecom wavelength range (Fig. 1.b). The front-end is carrier-envelope phase

(CEP) stabilized using an ultralow phase-noise feed-forward technique[20] that ensures single-digit mrad pulse-to-pulse jitter for practically long operation[21]. Stabilizing the CEP is the first step to guarantee pulse-train phase consistency across all beamlines. After splitting the CEP-stabilized front-end, one beamline sets a reference in order to monitor and control relative inter-beamline phase offset via a self-synchronous and self-referenced custom field-programmable gate array (FPGA) phase-locking technique. This configuration enables all 7 beamlines to be phase-locked to the absolute reference phase—which is set via CEP—by any arbitrary phase relationship. Thus, all but the reference beamline undergo active manipulation and monitoring of all the measurable field parameters—namely phase ($\Delta\phi_i$), amplitude ($A_i$), polarization state ($\wp_i$), and timing ($\Delta t_i$)—prior to their coherent synthesis or distributed delivery. Each beamline contains a phase modulator (here a piezoelectric transducer-based fiber stretcher) that imposes a user-defined phase relationship with respect to the reference beamline via computer-interfaced FPGA with a maximum programmable range of 20λ with these specific modulators. Active phase-locking provides 40 mrad phase noise, or 33 as of timing jitter, corresponding to an outstanding level of stability four orders of magnitude lower than the entire programmable range. The intensity and polarization vector control units for each beamline consist of a half waveplate, polarizing beam splitter, and quarter waveplate placed on a fiber pigtailed delay stage for timing. After individual manipulation of the field vectors, circularly birefringent fibers preserve each beamline's final polarization state prior to synthesis. The composite beam is collimated and synthesized in free space with a micro-lens array in a tiled-aperture configuration with the seven beamlines arranged hexagonally to be spatio-temporally overlapped at a photodiode. This photodiode is the only optical detection component required for the self-synchronous self-referenced locking technique. Further technical details can be found on Methods.

The resulting product is a self-consistent laser architecture that can deliver 4-D programmable pulses in the form of a free-space synthesized light bullet (e.g. near or far-field), as an array of distributed coherent beamlines (e.g. fiber), or as a hybrid distributed fiber- and free-space beamlines. To showcase some of these capabilities, Fig. 2 exemplifies the synthesis of various pulses in the far-field from combining only the amplitude and relative phase of the beamlines. Note that the near-field hexagonal arrangement is chosen here for demonstration purposes only. In order to highlight the effectiveness of these two knobs alone in generating complex intensity- and phase- distributions, we denote each beamline's relative phase difference with respect to the others ($\phi_k$) to be $\phi_k \leq 2\pi$ and for simplicity, the amplitude of each beamline ($A_k$) to be either 'on' or 'off', i.e. $A_k \in \{0,1\}$, where $k \in Z[1, N]$ and $N$ is the total number of beamlines in the combined output. The various $A_k$ and $\phi_k$ arrangements are displayed on column A in Fig. 2. Columns B and C show the corresponding calculated and measured far-field transverse intensity distributions, respectively. Column D is the corresponding retrieved phase distribution at the plane of the measurement. Row 1 exemplifies a conventional coherently combined fiber array, typically used for intensity scaling purposes. Rows

2–4 demonstrate cylindrically structured pulse synthesis with either alternating or gradually varying phase. For instance, in Row 3 the phase is defined as $\phi_k = 2\pi(k-1)/N$, that is, the total phase offset spans over $2\pi$ increasing clock-wise monotonically and uniformly, which is equivalent to generating a discretized first-order orbital angular momentum (OAM) beam. Note that while both the far-field intensity and phase resemble that which is expected, i.e. a Laguerre-Gauss and helical shapes, respectively, the additional structure in these distributions arises naturally from discretization. As the distributions in Row 2 and 3 are brought to the far field, a field singularity appears at the center of these beams. It is worth noting that this singularity shifts away from the center towards the three triangular-equidistant lobes in the case of Row 4. The direction of increasing phase in the near field can be switched to change the chirality of the OAM beam in the far field. Interestingly, despite the hexagonal arrangement, cylindrically asymmetric pulses can also be generated, as shown in Rows 5–8, some of which results in abrupt phase transitions with reflectional symmetry. Collectively, these showcases highlight the ability to create adaptive and dynamic field singularities and intensity distributions.

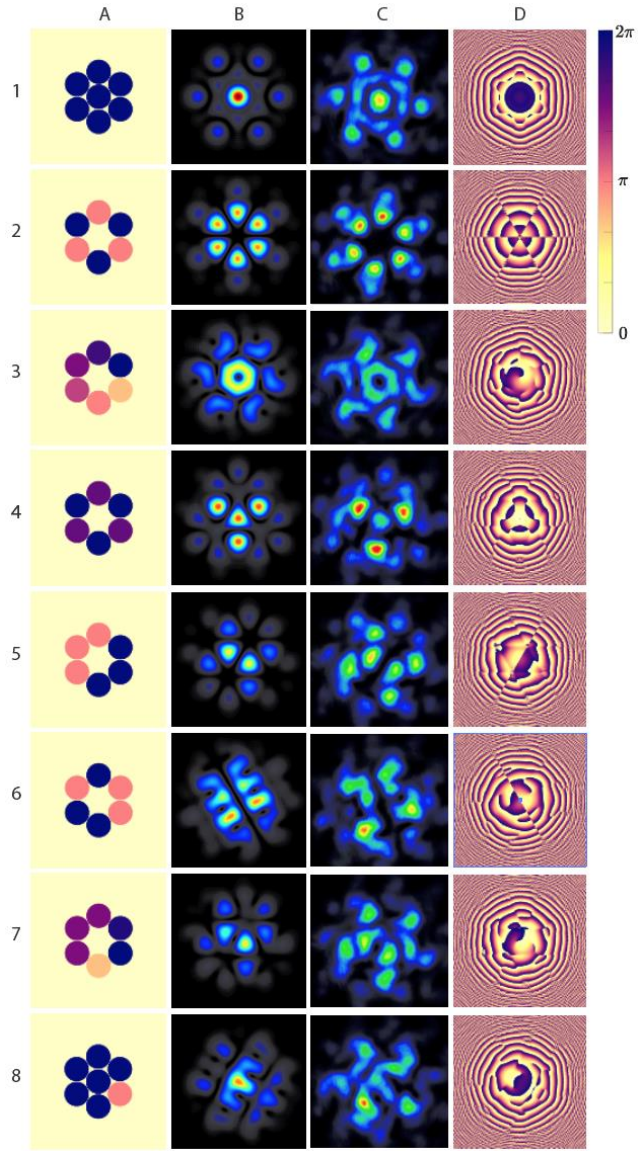

Fig. 2—Near-field phase- and amplitude combinations (A) and their corresponding retrieved (B) and measured synthesized far-field intensity (C) and phase distributions (D).

One of the most salient features of the ULM is the capacity to generate programmable composite phase-fronts in the near and far fields for a diverse array of composite polarization states, thereby producing beams with spatially and temporally variant spin angular momentum distributions. We highlight an evolution of a few possible topographic polarization distributions in Fig. 3.a–c, where each case represents the spin angular momentum distribution map overlaid on top of the far-field intensity distribution with the corresponding near-field configuration (top left) alongside the three measured Stokes' maps on the bottom. In these examples, we have chosen to maintain all beamlines set to approximately the same phase and amplitude value and focus on the non-uniform transverse polarization topography generated solely by varying the near-field polarization distribution. The vector maps evolve to follow constructive interference and a degree of ellipticity and chirality determined by the collective contribution of all beamlines at any transverse (x,y) point in the synthesized beam. Combining beamlines with orthogonal polarization states facilitates the generation of stable and adaptable interference patterns with alternating topological charge and singularity regions in the transition from one charge to another. The Stokes measured projections show further evidence of the transferability of the 2D topographic map to another state by highlighting the inverse relationship between the linear polarization projection map in (b) and the circular projection map in (c), that is, $S_1^b = -S_3^c$. Because the transverse polarization distribution arises from any combination of $\Delta\phi_i$, $A_i$, and $\wp_i$, an exceptionally large ensemble of topographic polarization maps is possible with only a few beamlines.

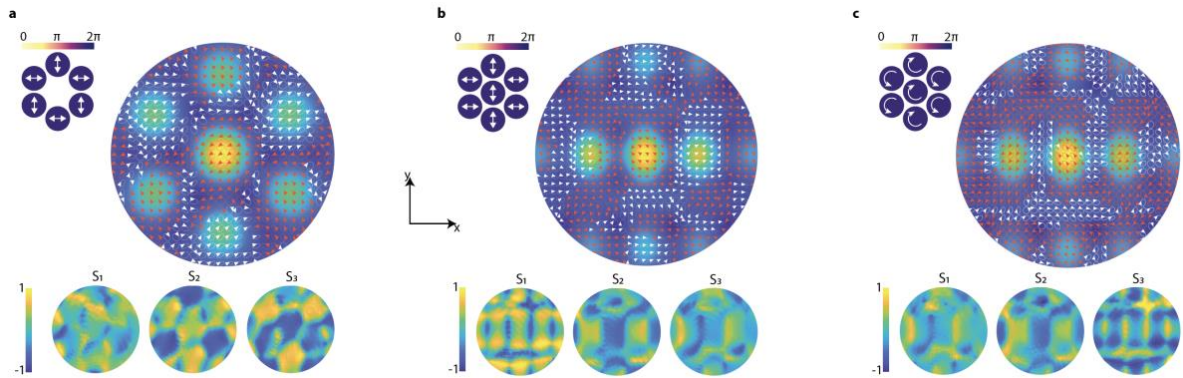

Fig. 3—polarization topography evolution with corresponding near-field configurations (top left) and Stokes projections (bottom) for alternating linear (a), asymmetric linear (b), and asymmetric circular polarization coherent synthesis (c).

One last consequential feature of this integrated structure light architecture resides entirely on the composite action of all $\Delta t_i$, which facilitate the generation of optical pistons from small delays (few to several wavelengths) with a time precision determined by the locking phase stability to very large delays beyond the duration of the pulses and dynamic range determined by optical delay stages. It is important to note that while all examples presented above are synthesized in the free space to highlight and diagnose the main architectural capabilities, beamline delivery need not only collapse in a single location but can also be configured as a distributed coherent array in any guided, unguided, or hybrid configuration.

This demonstration materializes a dynamical and programmable architecture to produce light by design, where spatio-temporal wavevector distributions can be tailored in real time to enable further exploration of structured photonics and its applications. In particular, the ULM showcases the synthesis of high geometrical dimensionalities of structure, including nontrivial vector map topologies in spatial and temporal dimensions. The system is self-encapsulated such that field controls are integrated in the architecture itself, here specifically providing femtosecond laser pulses with adaptive and precise synchronicity. It also presents a few notable imminent opportunities such as power-scalability and hyperspectral extension via integrated photonics to be examined in quantum electrodynamics and on-chip accelerator drivers[22]. More broadly, this novel architecture aspires to seed new frontiers of light control and manipulation, optical quantum communications and information processing, as well as emerging concepts in nonlinear topological and nuclear photonics.

### Acknowledgements
This letter was supported in part by the U.S. Department of Energy, Laboratory Directed Research and Development program at SLAC National Accelerator Laboratory under contract DE-AC02-76SF00515, and by the U.S. Department of Energy Office of Science under contract DE-SC0014664. We would also like to thank and acknowledge Greg Stewart for his artwork contributions.

### Author Contributions
S.C., J.R., A.F., and W.L. are the conceptual authors of the work. S.C., W.L., R.L., J.C.F. and S.S. designed and conducted experiments. All authors contributed to the manuscript.

### Competing Interests statement
No competing interests.

Methods

1. Carrier-envelope phase stabilized front-end and beamline controls

We use a soliton mode-locked Er:Yb:glass laser oscillator (OneFive Origami-15) and CEP stabilization based on a feed-forward (FF) system[21]. The oscillator delivers 140 mW of power in 175 fs pulses at a repetition rate of 204 MHz ($f_{REP}$) with a spectral bandwidth of 14.9 nm centered around 1.55 µm. The light from the oscillator is split into two beamlines: one towards the in-loop (IL) feedback measurement and the other through the acoustic-optic frequency shifter (AOFS) and towards the out-of-loop (OOL) measurement. Both beamlines are coupled into stretcher fiber, which is spliced with Er:fiber amplifiers. After nonlinear amplification the pulse is recompressed and passes through highly nonlinear fiber (HNLF) for octave spanning. The spectrally broadened pulses are coupled out to free space and frequency-doubled in a periodically-poled lithium niobite tuned for second harmonic generation at 1024 nm. The light is then passed through optical band pass filters centered at 1024 nm and focused on to an avalanche photo diode (APD).

The raw signal from the IL APD is sent to the FF electronics and conditioned for the AOFS, which has an operational frequency of 80 ± 2.5 MHz, since $f_{CEO}$ is not guaranteed to be in this range. The signal is filtered to isolate $f_{CEO}$ with 40 dB SNR (RBW:100 kHz), which is mixed with a local oscillator (LO) and amplified to 26 dBm. The final signal is given by $f_{AOFS} = f_{CEO} + f_{LO} = 80\ MHz$. The AOFS subtracts the drive signal from the frequency comb replacing $f_{CEO}$ with $f_{LO}$ and power is shifted to the AOFS -1st diffraction order and coupled into the OOL interferometer fiber. The raw signal measured in the OOL interferometer contains $f_{LO}, f_{REP}$, and mixing products. For jitter analysis, the signal is filtered to remove $f_{REP}$ and mixing products.

Long term operation is achieved by sending to $f_{AOFS}$ to a slow feedback loop that adjusts the pump power of the oscillator to correct for drifts in the system. The loop detects drift away from 80 MHz through mixing with a stable 79.8 MHz signal, a low pass filter at 500 kHz, and a frequency to voltage (F2V) converter. The F2V signal is then fed to a PID controller with only proportional and integral gains. The output of this PID is finally what drives the pump power changes in the oscillator.

We employ SPUN-HiBi fibers for individual beamline delivery. SPUN-HiBi are designed to preserve circular polarization. The composite beam is collimated with a microlens array in a tiled-aperture configuration arranged hexagonally. Relative time overlap/delay is achieved using a fiber pigtailed delay stage in each beamline. The intensity is modulated by a half waveplate and a polarizing beam splitter integrated in the delay line. Phase control is achieved by means of 7 phase modulators (PZT-based fiber stretcher) capable of operating at bandwidths larger than 10 kHz.

2. Multi-channel Phase Modulation: FPGA-based LOCSET

Optical phase control and modulation begin with alignment of the incoming optical phase of the seven channels, accomplished by overlapping all channels on a single photodiode (PD) and maximizing the amplitude seen by the diode. An optical phase error signal for each channel is generated by phase modulating (PM) each channel at a unique PM frequency of a few hundred Hz to tens of kHz. One beam may serve as a phase reference without modulation, though convergence is slightly faster if feedback is applied to all channels driving them all to the mean phase of the ensemble (referenceless operation). Front-end electronics down-convert the PD electrical signal from 204 MHz, corresponding to the nominal repetition rate of the laser, to 2 MHz and digitized at 20 MSPS by a 16-bit ADC whose output is streamed to the feedback FPGA. The FPGA digitally demodulates the 2 MHz IF then further demodulates each sideband at the PM frequencies of each of the seven channels. These sidebands may be coherently demodulated as the FPGA itself synthesizes the PM drive. The sign and amplitude of each sideband gives the optical phase error for each channel. Each beamlet phase error drives a feedback loop filter for that beamlet. The feedback bandwidth may be set as high as a few hundreds of Hz so phase convergence is attained quickly. Once optical coherence has been established at the photodiode, one can program the desired complex optical modulation program. As optical path lengths drift, optical phases must periodically realign by recohering the beams on the photodiode.

3. Beam Propagation Model

The free-space simulation of this system is a discrete fast Fourier transform angular spectrum evaluation method of the Rayleigh-Sommerfeld diffraction formula. The object plane for all simulations is placed at the micro-lens array and assumes that all beamlines are collimated. Additionally we place the image plane at our camera and assume that the propagation between the two planes satisfies all requirements for using a scalar rather than vector propagation theory[23]. The use of angular frequencies, and the resultant finite, discrete grid, introduces conditions on the proper sampling of the object and image planes, and their reciprocal spaces to obtain accurate results. First, all non-zero values of the objects must be included in the computational grid which is automatically satisfied if the extent of the grid is larger than the object[24]. Second, the maximum propagation distance without aliasing due to under sampling is related to the real space grid spacing, the wavelength of light, and the farthest extent in the grid which contains a non-zero value[25]. Third, the minimum propagation distance must be substantially greater than the wavelength[23]. Since our propagation is on the order of meters the third condition is satisfied and we simply increase the extent of the object grid to a large enough size that the second condition is met. Additionally, we model the full image plane of mixed polarization states at the object plane by propagating two fields, one for each orthogonal polarization, and taking the sum of their intensity distributions at the image plane.

## 4. Polarization Vector Map Calculations

The polarization vector maps are reconstructions of the local polarization ellipse generated from stokes parameters. In order to capture the Stokes parameters $\{S_0, S_1, S_2, S_3\}$, seven images, the full field and one for each of the six projections on the Poincaré sphere, are mixed according to the known definitions. Each projection image of size $\mathbb{N} \times \mathbb{M}$ is captured by system consisting of a quarter wave plate, a half wave plate, a polarizing beam splitter, and an InGaAs camera. To ensure that each projection image is capturing the same region of the field a smaller image of $n \times m$ pixels is taken from the full image and the normalized cross correlation between this and each projection is calculated. When the cross correlation is at its maximum the projection is assumed to be capturing the same region and then cropped to $N \times M$. Next to remove errors from shot to shot pixel differences the images are subdivided into $n \times m$ macro-pixels, where each macro-pixel contains the mean of a subset of true pixels $\alpha \times \beta$ such that $\alpha n = N$ and $\beta m = M$. After being centered and sub-divided, the seven images are finally used to calculate the local Stokes parameters of the field. To generate the polarization ellipse, it is necessary to know the eccentricity, the tilt relative to a fixed axis, and the chirality. The eccentricity is given by,

$$e = \sqrt{\frac{2\sqrt{S_1^2 + S_2^2}}{1 + \sqrt{S_1^2 + S_2^2}}} \qquad \text{(Equation 1)}$$

the tilt by,

$$2\theta = \tan^{-1}\frac{S_2}{S_1} \qquad \text{(Equation 2)}$$

and the chirality is determined from the sign of $S_3$. In Fig. 3 we reduced the number of plotted vectors to reduce visual clutter without eliminating the shape of the evolving field.